\documentclass[namedreferences]{Astrophysics}

\usepackage[hyperref,optionalrh]{spr-sola-addons} 
\usepackage{graphicx}        
\usepackage{multirow, threeparttable, booktabs,lscape}
\usepackage{color}           
\usepackage{breakurl}        

\usepackage{lmodern}%
\usepackage{anyfontsize}
\usepackage{geometry}

\usepackage{silence}

\usepackage{threeparttable}
\usepackage{longtable}
\usepackage{rotating}
\usepackage{pdflscape}
\usepackage{tabularx}
\usepackage{changepage}
\usepackage{amsthm, amssymb}

\usepackage[latin1]{inputenc}%
                     \usepackage[T1]{fontenc}%

\chardef\us=`\_

\begin{document}

\begin{article}
\begin{opening}

\title{Analysis of type II and type III radio bursts associated with SEPs from non-interacting/interacting radio-loud CMEs}

\author[addressref=aff1,corref,email={kalai2230@gmail.com}]{\fnm{P}~\lnm{Pappa Kalaivani}\orcid{0000-0002-7364-0768}}
\author[addressref={aff2,aff3}, email={prakash18941@gmail.com}]{\fnm{O}~\lnm{Prakash}\orcid{0000-0002-3312-0629}}
\author[addressref={aff4},email={ashanmugaraju@gmail.com}]{\fnm{A}~\lnm{Shanmugaraju}\orcid{0000-0002-2243-960X}}
\author[addressref={aff3},email={lfeng@pmo.ac.cn}]{\fnm{}~\lnm{Li Feng}\orcid{0000-0003-3636-2962}}
\author[addressref={aff3},email={leilu@pmo.ac.cn}]{\fnm{}~\lnm{Lei Lu}\orcid{0000-0002-3032-6066}}
\author[addressref={aff3},email={wqgan@pmo.ac.cn}]{\fnm{}~\lnm{Weiqun Gan}}
\author[addressref={aff5}, email={michalek@oa.uj.edu.pl}]{\fnm{G}~\lnm{Michalek}}

\address[id=aff1]{Department of Physics, Ultra College of Engineering and Technology for Women, Ultra Nagar, Madurai - 625 104, Tamil Nadu, India}
\address[id=aff2]{Department of Physics, Sethu Institute of Technology, Pulloor, Kariapatti,Viruthunagar, Tamil Nadu -- 626 115, India}
\address[id=aff4]{Department of Physics, Arul Anandar College, Karumathur, Madurai -625514, India}
\address[id=aff3]{Key Laboratory of Dark Matter and Space Astronomy, Purple Mountain Observatory, Chinese Academy of Sciences, Nanjing -- 210008, Jiangsu, China}
\address[id=aff5]{Astronomical Observatory of Jagiellonian University, Cracow, Poland}

\runningauthor{Pappa Kalaivani et al.}
\runningtitle{\textit{Analysis of type II and type III radio bursts}}

\begin{abstract}
We analyze radio bursts observed in events with interacting/non-interacting CMEs that produced major SEPs (Ip $>$ 10 MeV) fromApril 1997 to December 2014. We compare properties of meter (m), deca-hectometer (DH) type II as well as DH type III bursts, and time lags for interacting-CME-associated (IC) events and non-interacting-CME-associated (NIC) events. About 70\% of radio emissions were observed in events of both types from meters to kilometers. We found high correlations between the drift rates and mid-frequencies of type II radio bursts calculated as the mean geometric between their starting and ending frequencies for both NIC and IC-associated events (Correlation coefficient \textit{R}$^{2}$ = 0.98, power-law index $\varepsilon$ = 1.68 $\pm $ 0.16 and \textit{R}$^{2}$ = 0.93, $\varepsilon$ = 1.64 $\pm $ 0.19 respectively).We also found a correlation between the frequency drift rates of DH type II bursts and space speeds of CMEs in NIC-associated events. The absence of such correlation for IC-associated events confirms that the shock speeds changed in CME--CME interactions. For the events with western source locations, the mean peak intensity of SEPs in IC-associated events is four times larger than that in NIC-associated SEP events. From the mean time lags between the start times of SEP events and the start of m, DH type II, and DH type III radio bursts, we inferred that particle enhancements in NIC-associated SEP events occurred earlier than in IC-associated SEP events. The difference between NIC events and IC events in the mean values of parameters of type II and type III bursts is statistically insignificant.
  
\end{abstract}
\end{opening}

\section{Introduction}
     \label{S-Introduction} 

Coronal mass ejections (CMEs) are the biggest eruptive phenomena on the Sun. CMEs propagating in the solar corona and interplanetary space have an energy up to 1032 ergs and a mass up to 1016 kg [1, 2].The width of a CME is also significant for its kinetic energy [3, 4]. Full or partial halos surrounding fast CMEs are most likely shock signatures [5-7].Fast CMEs that exceed the local Alfvйn speed in the environment are capable of driving shocks in the coronal and interplanetary (IP) medium [8]. CME-driven shocks can accelerate protons and electrons to high energies. Accelerated electrons can excite plasma oscillations and generate type II bursts at the local plasma frequency and its harmonic.\\[-5pt]

A type II radio emission slowly drifts from higher to lower frequencies. The plasma frequency is directly proportional to the square root of the local plasma electron density. The meter (m) type II bursts are observed in a frequency range of 200-18 MHz by ground-based radio spectrographs. The deca-hectometer (DH) type II bursts are observed in a range of 14 MHz - 20 kHz by space-born radio instruments. The generation of type II radio bursts in various spectral domains depends on the kinetic energy of the associated CME. The so-called,  \lq \lq radio-loud\rq \rq CME are generally associated with type II radio bursts and, therefore, with interplanetary shocks [9].The lack of shock signatures for some fast and wide CMEs is still under debate.\\[-5pt]

Type III radio bursts are fast-drifting features caused by electron beams accelerated in solar flares [10-11]. Cane et al. [12] identified a new class of km type III bursts observed by theISEE-3 spacecraft. They found that some low-frequency type III bursts had unique properties such as a strong intensity and long duration when compared to normal type III bursts. This type of events was referred to as shock-associated (SA) events [13]. Later, Cane et al. [14] studied relationships between high-energy proton events, CMEs, solar flares, and radio bursts. They pointed out that these intense long-duration low-frequency type III bursts were caused by electron streams travelling from the Sun to 1 AU. They also reported that almost all intense long-duration low-frequency type III bursts were associated with CMEs. Having compared the type III intensity at 1 MHz in impulsive and gradual SEP events during solar cycle 23, Cliver and Ling [15]showed that almost 95\% of DH type II bursts were associated with gradual SEP events.Winter and Ledbetter [16] analysed 123 DH type II radio bursts observed by Wind/WAVES and associated SEP and non-SEP events during 2010-2013. They found that 92\% of SEP events were associated with both DH type II and DH type III bursts and proposed that the peak intensity and duration of DH type III bursts were the dominant factors among the properties of radio bursts. They concluded that DH type III bursts accompanying DH type II bursts could be used to forecast SEP events. Pappa Kalaivani et al.[17] also found for a set of events a closer association of peak intensities of SEPs with properties of DH type III bursts than with properties of DH type II bursts.\\[-5pt]

Pappa Kalaivani et al. [18] (hereinafter Paper I) suggested that CME?CME interactions could enhance the peak intensity of SEPs. The seed populations in interacting-CME-associated (IC) events and non-interacting-CME-associated (NIC) events were found to be similar. A number of case studies reported that after an interaction between CMEs the intensity of DH type II bursts increased [3, 19, 20]. However, there is no comprehensive analysis of (i) properties of type II and type III bursts that accompany IC events and (ii) their comparison with NIC-associated events. The main objective of this study is to find if any differences exist between properties of type II (m and DH) and DH type III bursts accompanying NIC and IC events. We present the event selection and analysis in Section 2 and discuss the results in Section 3. Section 4 presents a summary and conclusions.

\section{Event selection and Analysis}
We have utilized 143 major SEP events (\textit{I}$_{p}$ $>$ 10 MeV) that occurred between November 1997 and December 2014 from the catalog compiled by the Coordinated Data Analysis Workshops (CDAW) data center\footnote{$https://cdaw.gsfc.nasa.gov/CME_list/sepe$}. A set of 125 major SEPs is considered by excluding 18 ground level enhancement (GLE) events, as done in Paper I. Out of these 125 events, we identified 70 major SEPs using simple selection criteria as follows: i) the data on solar flares and CMEs must be unambiguous; the flare locations must be clear and more than three height-time (h-t) data points must be observed for a CME, and ii)flare-associated CMEs must be associated with both m-and DH type II radio bursts. From type II catalogues we have compiled the following data set: i)the end time, start frequency, end frequency of m type II radio bursts, and estimated shock speeds reported by several ground-based spectrographs during 1994--2009were taken from the online catalog\footnote{$https://www.ngdc.noaa.gov/stp/space-weather/solar-data/solar-features/solar-radio/radio bursts/reports/spectral-listings/Type_II/Type_II_1994-2009$} that is maintained by the National Geophysical Data Center (NGDC) and monitored by the National Oceanic and Atmospheric Administration (NOAA). In addition, we compiled the data on m type II bursts observed during 2010 -- 2014from dynamic spectra provided by the Radio Solar Telescope Network (RSTN)\footnote{$https://www.ngdc.noaa.gov/stp/space-weather/solar-data/solar-features/solar-radio/rstn-spectral$}; ii) the end time, start and end frequencies of DH type II bursts were taken from the Wind/WAVES website\footnote{$http://ssed.gsfc.nasa.gov/waves/data_products.html$}. The association of major SEPs of m and DH type II bursts was adopted from the CDAW\rq s major SEP event list.\\[-5pt]

To select shock properties, we preferentially consider the fundamental emission of each m type II burst, if it has both fundamental and harmonic emissions. For this set of 70 pairs of m-and DH type II bursts, a log frequency-time plot (drift graph) is drawn for each event using the start and end frequencies and times of both m and DH type II bursts using the method described by Prakash et al. [21]. Using the drift graphs, we composed a set of 58 DH type II radio bursts in Paper I as continuation of m type II bursts (hereinafter m-to-DH type II bursts) for our further analysis. Furthermore, we have divided these 58 major SEP-associated m-to-DH type II burst events into two categories by checking if a SEP-associated CME(primary CME) interacts with a slow preceding CME(IC event) or not (NIC event). We verified the CME interaction from the height?time (h?t) data in the field of view (FOV) of the Large Angle and Spectrometric Coronagraph (LASCO) on board the Solar and Heliospheric Observatory (SOHO), and the overlap of CMEs in the position angle (PA) in SOHO/LASCO movies. In addition, we checked whether there are significant changes in the kinematics of both preceding and primary CMEs from the h-t data. Due to the data availability of the Solar Terrestrial Relations Observatory (STEREO) from late 2006, almost one-third of CME interactions were further confirmed from different vantage points by the overlap in PA in the movies of the COR2 coronagraph on board STEREO. 

\subsection{Analysis of DH type II and DH type III radio bursts}

This subsection describes how the basic properties of DH type II and DH type III radio bursts (\emph{i.e.,} peak intensity, integrated intensity, slope, and duration) are derived from dynamic radio spectra. To derive the properties of the present set of 58 events, we adopted the method of Winter and Ledbetter [16]. In general, DH type II and DH type III bursts are recorded by three detectors of the Wind/WAVES instrument, RAD1 (20 -- 1040 kHz), RAD2 (1.075 -- 13.825 MHz) and TNR (4 -- 245 kHz) [22-23]. The calibrated one-minute average data from these three detectors were downloaded as IDL save files for each event. In each file, the ratio (R) to the background values (B) in the units of $\mu$V Hz$^{-1/2}$ is recorded. We converted the R and B values to solar flux units (1 sfu = 10$^{-22}$ W m$^{-2}$ Hz$^{-1}$) by using the relation J (sfu) = 10$^{10}$ (R $\times$ B) / (Z$_{o}$ $\times$ A), where A is the area of the antenna, Z$_{o}$ is the impedance of free space. Additional corrections were also made.

\begin{figure} [ht]   
   \centerline{\includegraphics[width=1.\textwidth,clip=]{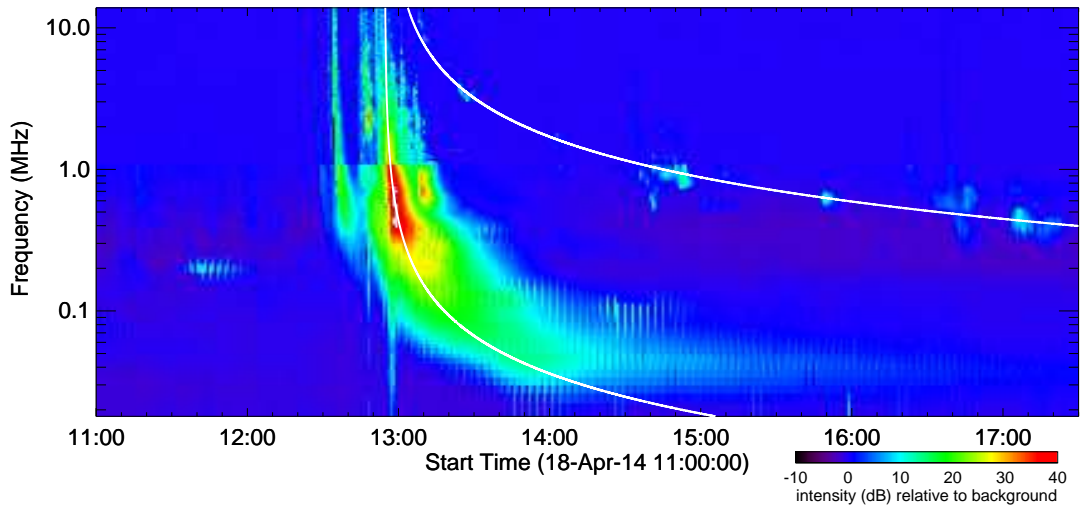}
              }
              \caption{Example of a dynamic spectrum of DH type II and DH type III bursts.The white solid lines represent the linearly fitted curve to the local peak intensities at each frequency.}
   \end{figure}

For a given DH type II burst, we plotted the dynamic spectra and picked up the local peak intensity at each frequency. The corresponding frequency and time data were recorded. A linear function was then fitted to the 1/frequency and time points to yield a slope (frequency drift). The result of the linear fit was transformed back to the frequency-time space to trace the type II burst, which is shown by the upper white curve in Figure 1. We estimated the slope and the integrated intensity by integrating along the fitted line both in time and frequency between the points where the flux decreased to 15\% of the local peak intensity.\\[-5pt]

A similar method was used to estimate the peak intensity, integrated intensity, and the slope of a complex DH type III radio burst. The time span when the signal at 1 MHz exceeded 6 dB (fourfold the background value) [24, 15] was taken as the duration of the burst. Table 2 (see Appendix) lists the derived parameters of DH type II bursts, DH type III bursts, and m type II bursts for completeness. The parameters for IC and NIC events are presented separately. Column 1 given the events number. The start date of a SEP event is given in Column 2. Columns 3--7 list parameters of m type II bursts. Columns 8--13 list parameters of DH type II bursts. Columns 14--16 list parameters of DH type III bursts.


\subsection{Statistical analysis methods}

Table 1 presents the mean and the standard deviation ($\sigma$) of the parameters forthe m, DH type II, DH type III radio bursts, and the time lags found separately for NIC and IC events. Column 1 gives the name of the object. Column 2 denotes the properties of the considered events. Columns 3, 4 (NIC) and 5, 6 (IC) list the estimated mean and standard deviation of the respective parameters. In Column 7, the statistical significance of the null hypothesis (P value) from the Student\rq s \emph{t} - test is given in percent. The Student\rq s \emph{t}-distribution has a higher kurtosis than normal distributions. It has a higher probability for extreme values on either side of the distribution peak. This is due to outliers which are observed in extreme cases or due to variability in measurements and experimental errors. The outlier is an extreme value, being any data point that lies over \emph{1.5 IQR} (inter quartile range) below the first quartile (\emph{Q1}) or above the third quartile (\emph{Q3}) in a data set. The inter quartile range is a measure of the statistical dispersion. It is the middle 50\% or the so-called H-spread. It is obtained by the first quartile subtracted from the third quartile. If the P value is less than 5\%, then there is a 5\% probability that the difference is accidental and that the statistical significance is 95\%. In the next section, we present the results of the statistical analysis and discussion over the IC and NIC events. 

\begin{table}[htp]
\caption{Mean and standard deviation ($\sigma$) of different properties of m type II bursts, DH type II bursts, and DH type III burst for NIC and IC-associated events.}
\begin{center}
\begin{tabular}{|c|l|c|c|c|c|c|}
\hline

\multirow{2}{*}{Events} &\multirow{2}{*}{Properties}& \multicolumn{2}{c}{NIC events}& \multicolumn{2}{c} {IC events} & \multirow{2}{*}{P value}\\
& &Mean & $\sigma$ & Mean & $\sigma$ &  \\\hline 

\multirow{4}{*}{m type II} & Duration (min) & 11 & 6 & 9 & 5 & 32\%\\
&Starting frequency (MHz) & 101 & 47 & 108 & 58 & 61\%\\
&End frequency (MHz)& 25 & 6& 28 &  16& 26\%\\
&Estimated shock  speed (km s$^{-1}$)& 957 & 427 & 1110 &  407 & 18\%\\\hline

\multirow{6}{*}{DH type II} & Duration (min)$^{@}$ & 412  & 834& 925& 847& 37\%\\

&Starting frequency (kHz)
&12100
&4200
&11500
&4600
&64\%\\
&Ending frequency (kHz)$^{@}$
&1043
&2187
&400
&550
&17\%\\
&Peak intensity (sfu)
&2 $\times$10$^{4}$
&3 $\times$10$^{4}$
&1$\times$10$^{5}$
&5$\times$10$^{5}$
&22\%\\
&Integrated intensity (sfu Hz min)
&5$\times$10$^{7}$
&8$\times$10$^{7}$
&6$\times$10$^{7}$
&1$\times$10$^{8}$
&55\%\\
&Slope (MHz h$^{-1}$)
&0.67
&0.37
&0.75
&0.38
&43\%\\\hline

\multirow{4}{*}{DH type III}
&Duration (min)
&30
&10
&31
&12
&93\%\\
&peak intensity (sfu)
&4$\times$10$^{6}$
&5 $\times$10$^{6}$
&7$\times$10$^{6}$
&1$\times$10$^{7}$
&14\%\\
&Integrated intensity (sfu Hz min)
&2 $\times$10$^{10}$
&2 $\times$10$^{10}$
&3 $\times$10$^{10}$
&3 $\times$10$^{10}$
&18\%\\
&Slope (MHz h$^{-1}$)
&21
&7
&20
&8
&61\%\\\hline

\multirow{5}{*}{Time Lags}

&CME onset$^{*}$ -- SEP start delay (min)$^{@}$
&73
&38
&88
&46
&31\%\\
&Flare onset -- SEP start delay (min)$^{@}$
&78
&38
&96
&51
&23\%\\
&m type II -- SEP start delay (min)$^{@}$
&65
&39
&77
&46
&44\%\\
&DH type II -- SEP start delay (min)$^{@}$
&51
&41
&65
&50
&38\%\\
&DH type III -- SEP start delay (min)
&56
&47
&72
&43
&37\%\\\hline

\end{tabular}
\begin{tablenotes}
\small
\item The time lag analysis is presented only for NIC and IC events whose sources were located in the western hemisphere (L $\geq$ W30$^{o}$). Notations in Column 2: $^{*}$means that the CME onset time was estimated from the quadratic back-extrapolation to 1 \emph{R}$_{o}$ in the height -- time plot. $^{@}$These values are sensitive to outliers (see Section 3.2).
\end{tablenotes}
\end{center}
\end{table}%

\section{Results and Discussion}
\subsection{Properties of m type II radio bursts}
In addition to the properties of parent solar eruptions (Paper I), here we analyze the properties of m-to-DH type II and DH type III radio bursts for IC and NIC events. Our estimates are consistent with previous studies [26--27]. There is no significant difference between the mean durations of NIC and IC-associated m type II bursts. For both sets of events, the starting frequencies (\emph{f}$_{s}$) range from 200 MHz to 30 MHz with a standard deviation ($\sigma$) of 50 MHz. The mean starting frequencies in NIC and IC events are similar, being slightly higher than the value of 86 MHz found by Prakash et al. (2017) for m type II radio bursts associated with SEPs. We also found no significant differences between the average ending frequencies (\emph{f}$_{e}$) of the m type II bursts in NIC and IC events. Table 2 does not contain the estimated shock speeds for events 3 and 41, which are not reported in the online type II catalog, and for event 22, where the radio emission structure was too complex. The Student t-test shows that the difference between the NIC and IC events in the mean parameters of m type II bursts and the estimated shock speeds is statistically insignificant (see Table 2). Hence, he CME--CME interaction does not alter the properties of m type II radio bursts. This can be expected, because the projected interaction heights of CMEs mostly exceed 4 R$_{o}$ as discussed in Section 3.6.\\[-5pt]

\subsection{Properties of DH type II radio bursts}

In this subsection, we present the differences in basic characteristics of DH type II and DH type III radio bursts for IC and NIC-associated events. The DH type II radio bursts are more closely associated with fast and wide CMEs. Furthermore, m-to-DH type II radio bursts also replicate the energy of the associated CMEs [28, 8].\\[-5pt]

The starting frequency of DH type II bursts does not affect both sets of events because 62\% of DH type II radio bursts were observed with an upper cut-off frequency by WAVES instruments (14 MHz for Wind and 16 MHz for SWAVES). Hence, the difference between the mean starting frequencies of DH type II bursts for NIC and IC-associated events is statistically insignificant. For the DH type II durations, we obtained a mean value of 819 min with $\sigma$ = 834 min for NIC events and a mean value of 1036 min with $\sigma$ = 983 min for IC events. The significance of this difference shown by the student?s t-test is due to a single outlier (\#44) with an exceptionally long duration. Without this outlier, the mean duration for IC events is 925 min with $\sigma$ = 847 min and the difference with NIC events becomes insignificant. Because a typical event cannot determine the properties of the whole set, we conclude that the difference between the durations of DH type II bursts in the two categories of events is not significant. We also found that 69\% and 74\% of NIC and IC-associated DH type II bursts ending frequencies were observed at below 300 kHz (km domain). The difference between the mean ending frequencies for NIC and IC-associated events is statistically significant (P = 5\%). After removing the outliers from NIC events (\#3, \#6, \#19, \#24, and \#33) and IC-associated (\#40, \#45, \#48, and \#52) events, the mean difference  (293 and 169 kHz, respectively) is also statistically insignificant (P = 20\%). The outliers were noted from the distribution plots not shown here.\\[-5pt]

For NIC events, the outliers associated with DH type II bursts (\#3, \#6, \#18, \#24, and \#33) have short durations (15, 15, 35, 15, and 24 min, respectively) than the other events observed in the DH domain, and also, these outliers? associated SEP events have lower peak intensities (55, 12, 82, 26, and 22 pfu, respectively). These outliers? sources are located near the west limb (longitude $>$ W50$^{o}$). Furthermore, among the 5 outliers, \#6 was associated with the slowest CME (478 km s$^{-1}$). It is possible that the DH type II in this event had a short duration, likely because the shock would have rapidly ceased. But \#3 \#18, \#24, and \#33 were related to fast CMEs, whose speeds exceeded $>$ 900 km s$^{-1}$. Similarly, for IC events, these outliers? (\#40, \#45, \#48, and \#52) associated DH type II bursts have short durations than other IC events. These DH type II bursts associated CMEs are related to fast CMEs, whose speeds are larger than 1000 km s$^{-1}$. The peak intensities of these events are 15, 28, 466, and 215 pfu, respectively.  The source locations of these events are on the western side expect \#45.  Note that the presence of type II emission is considered as evidence of a shock, but the absence of type II emission does not guarantee the absence of shock. The short durations of these DH type IIs may be caused by emission propagation effects or unfavorable conditions for the generation of radio emissions.\\[-5pt]

We derived the peak intensity, integrated intensity, and slope of DH type II radio bursts for NIC and IC-associated events. On average, the peak intensity of DH type II bursts associated with IC is one order larger than that of NIC-associated events (see in Table 2).However, after excluding the outliers the mean peak intensities of DH type II bursts for NIC and IC-associated events (9 $\times$ 10$^{3}$ $\pm$ 1 $\times$ 10$^{4}$ sfu and 5 $\times$ 10$^{3}$ $\pm$  6 $\times$ 10$^{3}$ sfu, respectively) are statistically insignificant. For 74\% and 81\% of the IC-associated DH type II bursts the peak intensity is larger than 1000 sfu. The mean integrated intensity of DH type II bursts for IC-associated events is slightly larger than that of NIC-associated events.  Interestingly, 52\% of NIC-associated DH type II bursts have integrated intensity below 107 sfu Hz min. But for IC-associated events, only 32\% of DH type II bursts have integrated intensity less than 107 sfu Hz min. However, the difference between the means rejects the null hypothesis (P = 55\%). The slope of DH radio bursts (type II and type III) indicates the nature of drift rate and shock speed of associated shock in the interplanetary medium. The slope of DH type II bursts varies from 0.16 MHz h$^{-1}$ to 1.91 MHz h$^{-1}$. The slopes of IC-associated DH type II bursts are slightly larger than the NIC-associated events. Mean difference also rejects null hypothesis (P = 43\%).\\[-5pt]

\begin{figure} [ht]   
 \centerline{\includegraphics[width=1.\textwidth,clip=]{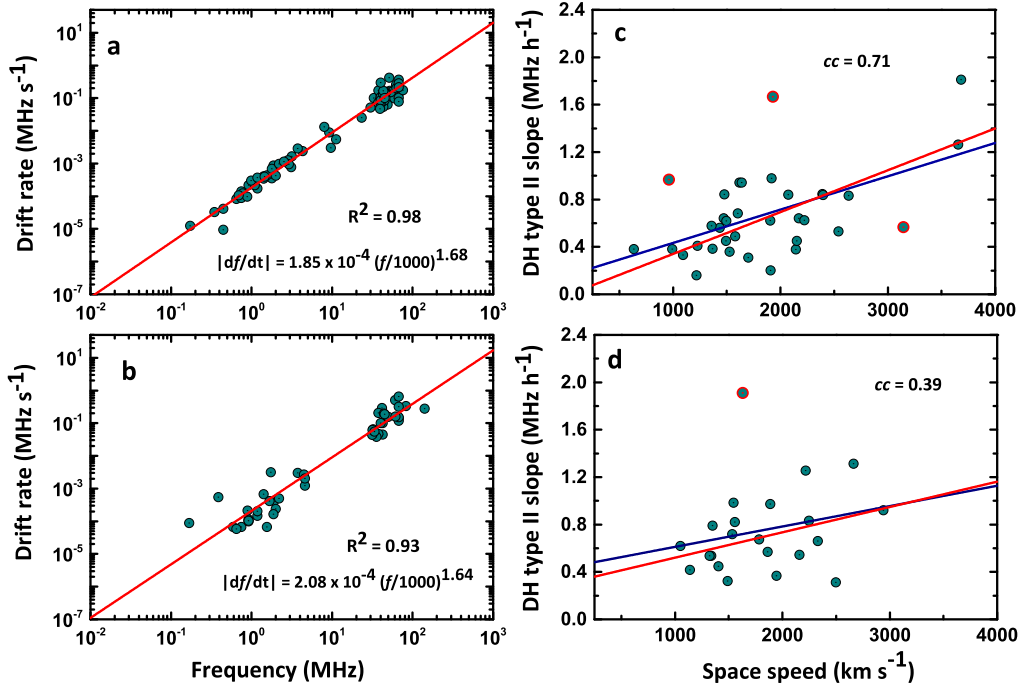}}
  \caption{Correlation between the geometric mean frequencies and drift rates of m and DH type II radio bursts for (a) NIC and (b) IC-associated events. The data points are fitted using the power-law in the form of $|df/dt|$= A\emph{f}$^{\varepsilon}$. Correlation between the slope of DH type II bursts and space speed of the CME for (c) NIC and (d) IC-associated events. Linear fits are shown before and after the exclusion of the outliers (red circled data points) by blue and red straight lines, respectively.}
\end{figure}

We studied the frequency dependence of the drift rates for m and DH type II radio bursts. The drift rate is estimated from the starting and ending frequencies, and the duration of a type II burst. It is important that the drift rate calculated in this way corresponds to the slope of the secant line of the type II trajectory at a mid-frequency and not to the slope of the tangent line at the starting frequency. We calculated the mid-frequency as the geometric mean between the starting and ending frequencies $f= \sqrt{f_{s}\times f_{e}}$ of each type II burst. As seen in Figure 2a--b, there are distinct correlations between the drift rates and the mid-frequencies of type II radio bursts for NIC and IC-associated events (\emph{R}$^{2}$= 0.98 and \emph{R}$^{2}$= 0.93, respectively) with power-law indices of $\varepsilon$ = 1.68 $\pm$ 0.16 and $\varepsilon$ = 1.64 $\pm$ 0.19, respectively. These correlation coefficients are considerably higher than 0.83 that was obtained by Umuhire et al [29] for a set of events. The data points in the metric range have a wider scatter relative to the DH range and to firm overall trends. This overall trend is possible due to the consideration of radio bursts whose trajectories continue from highest to lowest frequencies (m-to-DH type II bursts) without jumps. Also, the reduced drift rate in the DH range in Figure 2b may possibly be due to the two lowest-frequency outliers. If we exclude these two outliers from the power-law fitting, then the correlation coefficient becomes 0.96 with an increased index of 1.75 $\pm$ 0.18. For a combined data set of NIC and IC-associated events, we have obtained the frequency dependence for the drift rate with a power-law index of 1.66 $\pm$ 0.21.\\[-5pt]

The speed of a CME may be correlated with a drift rate of a type II radio burst. As we mentioned earlier in Sec.2.1, the slope (frequency drift) of the DH type II bursts were estimated when the data are transformed into 1/frequency space that are linear in time [30].Using this dependence, we can refer to all of the slopes of DH type II bursts for a geometrical mean frequency to 5 MHz for a representative sample of a set of events with starting frequencies ranging from 16 to 2 MHz. The relationship between the slope of the DH type II bursts and the space speed of CMEs is plotted in Figure 2c and 2d for NIC and IC-associated events, respectively. It is interesting to note that a positive correlation for NIC-associated events for these properties (\emph{cc} = 0.54), and the Pearson?s correlation coefficient increased as \emph{cc} = 0.71 after excluding the outliers. But, we found that there is no clear correlation between the slope of the DH type II bursts and the speed of CMEs for IC-associated events (\emph{cc} = 0.39, after excluding an outlier).The absence of correlation for DH type II bursts? properties and speed of CMEs for IC-associated events might be due to the changes in shock speeds in the interplanetary medium following the CME--CME interactions.

\subsection{Properties of DH type III radio bursts}

As we have discussed earlier, many authors pointed out the importance of complex DH type III radio bursts accompanying the DH type II bursts in relation with SEP events. Hence the properties of DH type III bursts are also studied in the present paper. Some of the authors studied the duration of complex DH type III bursts observed at 14 MHz and 1MHz signals from the radio dynamic spectra. In our case, we determined the duration of complex DH type III bursts at 1 MHz signal where DH type III intensity exceeds 6 dB or four times of the background as discussed in MacDowall et al. [24]. We are much aware of the duration of type III burst as responsible for the electron acceleration from the reconnection site. But the long duration of complex DH type III bursts at 1 MHz signal may also be contributed by the associated CME-driven shocks [12, 31].\\[-5pt]

The duration of NIC-associated DH type III bursts ranges from 11 to 65 min. But for IC-associated events, it lies between 5 and 48 min. The shortest complex DH type III duration at 1 MHz was on 2004 July 25 (5 min). Almost 51\% and 65\% of the durations of DH type III bursts are larger than 30 min for NIC and IC-associated events, respectively. However, the mean duration of DH type III bursts is similar for both sets of events. Most (89\%) of the complex DH type III bursts have a peak intensity below 1 MHz range. The intensity distribution is almost symmetric on either side of the peak value for both sets of events. Winter and Ledbetter [16] reported a mean peak intensity (1 $\times$10$^{7}$ sfu) of a different set of events which is larger than the mean peak intensity of DH type III bursts of all 58 events considered here. However, the mean difference between the peak intensity of the DH type III radio bursts of NIC and IC-associated events is statistically insignificant. We also derived the integrated intensity of complex DH type III radio bursts and the mean value of NIC-associated events is found to be only slightly lower than that of IC-associated events. On average, the slope of complex DH type III radio bursts of NIC-associated events is similar to the slope of the IC-associated events.

\subsection{SEP dependency on source locations and time lags studies}

The SEP propagation in the interplanetary space along the Parker spiral determines reduced peak fluxes and longer rise times of SEPs from eastern solar sources. The dependence of the SEP peak flux on the longitude of its source has been established [32, 33, 34, 35], but we are not aware of any expression of the longitudinal dependence of temporal parameters of SEPs. Hence, we divided our sample of events based on the longitudes (L): i) eastern side (L $>$ E30$^{o}$); ii) disk center (E30$^{o}$ $<$ L $<$ W30$^{o}$); and iii) western side (L $>$ W30$^{o}$). Figure 3 shows the distributions of the source location for all major SEPs, NIC-associated SEPs, and IC-associated SEP events.\\[-5pt]

Figure 3a gives the distribution of all 104 major SEP events during 1997--2014 (after excluding the backside events). Out of 104 events, 14\% and 59\% of SEP-associated sources are located on the eastern and western side of the solar disk, respectively and only 27\% of events are from the disk center. The distributions of the source locations for NIC and IC events are shown in Figures 3b and 3c, respectively. The numbers of events from eastern and disk center sources in the NIC and IC categories are insufficient for their separate statistical analysis. We, therefore, consider only those events, whose sources were located on the western side, and analyze SEP peak fluxes and their temporal parameters without any longitudinal corrections.

\begin{figure} [ht]   
 \centerline{\includegraphics[width=1.\textwidth,clip=]{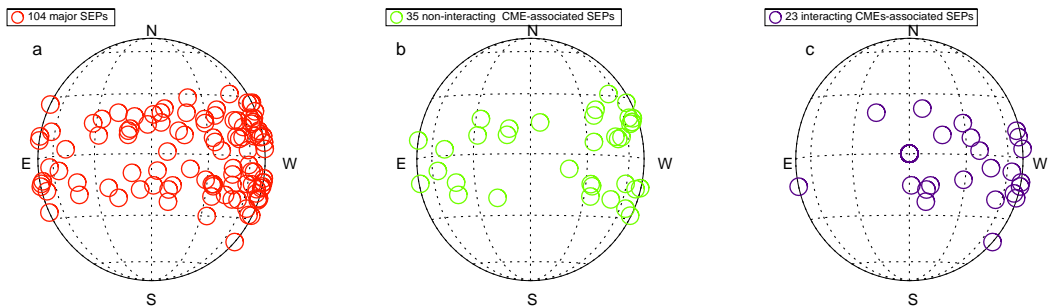}}
  \caption{Source distributions of (a) all major SEPs (b) NIC-associated SEPs and (c) IC-associated  SEP events.}
\end{figure}

Almost 63\% and 57\% of NIC and IC-associated SEP events, originated from the western side events.  The rise time and duration of the NIC-associated SEP events (570 min and 3 days, respectively) are slightly lower than the IC-associated events (635 min and 4 days, respectively). The distribution of peak intensity of the SEP events (which are from western side events) is presented in Figure 4. We noted that the distributions of the SEP intensity are similar for NIC and IC-associated events below 1000 pfu, while there is an excess of IC-associated events above 1000 pfu, although their number is meager. The peak intensity of SEPs for IC-associated events (1654 pfu) is three times larger than that NIC-associated SEP events (492 pfu). But after excluding the outliers from NIC (\#5, \#10, \#12) and IC-associated (\#41) events the mean peak intensity of SEPs for IC-associated events (559 pfu) is four times larger than that of NIC-associated SEP events (126 pfu). It is inferred that CME--CME interaction has played a vital role in the enhancement of the peak intensity of the SEP events. But the mean sky-plane speeds of the CMEs for both sets of events are similar.\\[-5pt]

\begin{figure} [ht]   
 \centerline{\includegraphics[width=0.7 \columnwidth]{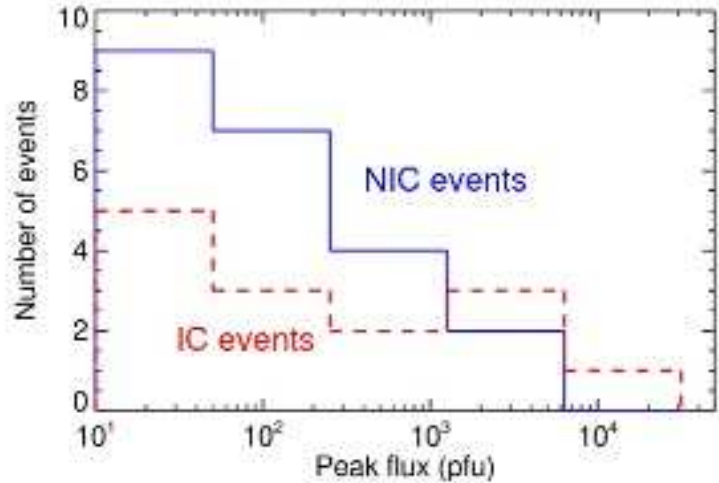}}
  \caption{Distribution of peak flux intensity of NIC and IC-associated SEP events with western solar sources.}
\end{figure}

The study of temporal association is also important to predict the occurrence of SEP events from their associated parent solar eruptions. Hence, we examined the various relative timings (time lags) between the start of SEP events and onsets of CMEs, solar flares, m, DH type II, and DH type III radio bursts for both sets of events. As we mentioned earlier, we have considered only those events from the western side events. The CME onset times were adapted from the quadratic back extrapolation to the solar surface as in Paper I. The time lag of CME onset from the start of SEPs for NIC-associated event ranges from 24 min to 173 min (after excluding an outlier \#11). But the distribution of time lags for IC-associated events is almost similar to NIC-associated events with a slight large $\sigma$ = 46 min. The mean time lag for NIC-associated events is less than that of IC-associated events. The Student t-test shows that the difference in the mean time lag is statistically insignificant (P = 31\%). For 48\% of NIC events, particle enhancement occurred within 60 min. But for IC-associated events, only 30\% events started within 60 min. We also observed that a similar kind of distributions for the other time lags. The time lag between the onset of solar flares and SEPs for NIC-associated events is lower than that of IC-associated events. The mean difference between NIC and IC-associated events is statistically insignificant.\\[-5pt]

We have also considered the time lags on the start of type II radio emissions (m- and DH type IIs) and the onset time of SEP events. The average time lag between the start of m type II/DH type II bursts and the onset of SEP events for NIC-associated events is also lower than that of IC-associated events and, however, the difference in the mean is statistically insignificant. From these time lag studies, we observed that NIC-associated SEPs started to enhance at least 10 min earlier than those in the IC-associated SEP events. 

\subsection{Characteristics of preceding CMEs}

To understand the reasons for small deviations in the derived properties of DH type II bursts between IC and NIC-associated events, we also analyzed the characteristics of preceding CMEs in IC-associated events. Mean sky-plane speed and mean width of preceding CMEs are 505 km s$^{-1}$ and 68$^{o}$, respectively.  The sky-plane speeds and widths of preceding CMEs are respectively three and five times lower than that of primary CMEs (see Paper I). Out of 23 events, 20 preceding CMEs (87\%) have widths less than 100? and the remaining three events (\#51, \#54, and \#57) have widths larger than 100$^{o}$. Also, 87\% of the preceding CMEs have a sky-plane speed less than 900 km s$^{-1}$ and only three events (\#37, \#38, and \#45) have sky-plane speed larger than 900 km s$^{-1}$ (1040 km s$^{-1}$, 929 km s$^{-1}$, and 1052 km s$^{-1}$, respectively). Also, 30\% of preceding CMEs were reported as poor for observation and one event (\#46) was reported as a streamer and its associated peak intensity of SEP event is 404 pfu. For each event, we also considered the average speed of preceding CMEs at 20 \emph{R}$_{o}$. We found that there are significant changes in the kinematic properties of preceding CMEs from the h--t data. An average (median) speed of the preceding CME is considerably increased as 724 km s$^{-1}$ (625 km s$^{-1}$) at 20 \emph{R}$_{o}$. It implies that slow and narrow preceding CMEs gain some kinetic energy from the faster primary CMEs upon interaction. It is in favor of the possibility of CME-CME interactions. It is also found that most of the preceding CMEs are accelerated within the LASCO FOV except for one decelerated event (\#37).\\[-5pt]

It should be noted that CMEs? interaction time and height were estimated when/where the leading edge of the preceding and primary CMEs get coincidence (Note that h-t error was unavoidable due to projection effect for disk CMEs). The projected interaction heights are widely distributed over the range between 3 \emph{R}$_{o}$ and 28 \emph{R}$_{o}$. An average projected interaction height is 13 $\pm$ 3 \emph{R}$_{o}$. It is lower than the mean projected interaction height for a set of SEP-associated events (21 \emph{R}$_{o}$) observed by Gopalswamy et al. [3].\\[-5pt]

We examined the time lags between the interaction time and start of SEPs starting time of DH type II bursts of IC-associated events. The time lags are distributed from --129 min to 339 min.  The particle enhancement started already before the interaction of CMEs for six events (26\%), and the remaining proton flux events started to increase after the interaction of the primary and preceding CME. We found a mean time lag of 63 $\pm$ 52 min. It implies that the shock interaction has been already progressive in the interplanetary space before the interaction of the leading edge of the CMEs. From these results, it may be concluded that the particle acceleration was already started and CME--CME interaction helps to enhance the peak intensity of the SEP events. Also preceding CMEs provide a significant number of seed particles to enhance particles acceleration much earlier than in the case of NIC-associated events.

\section{Summary and Conclusions}

In Paper I [18], we compiled a set of 58 m-to-DH type II radio bursts (m-to-DH: DH type II burst is a continuation of m type II burst) associated with major SEPs (\emph{I}$_{p}$ $>$ 10 MeV) and solar flares for IC and NIC events during the period April 1997--December 2014. The results obtained from the analysis of properties of solar flares, CMEs, and SEPs were presented in Paper I. In the present paper, we have listed a complete data set and comprehensively compared the properties of m, DH type II, and DH type III bursts for both sets of IC and NIC-associated events. The results of the present analysis of characteristics of m, DH type II bursts, DH type III bursts, and time lags for NIC and IC-associated events can be summarized as follows.\\[-5pt]

The duration and starting frequency of m type II burst are found to be almost similar for NIC and IC-associated events. The shock speed of m type II bursts is slightly larger for IC-associated events than for the NIC-associated events. We found that 69\% and 74\% of NIC and IC-associated radio emissions respectively were observed in the m-to-km type II bursts. The difference in the mean end frequencies between IC and NIC-associated events (400 kHz and 1043 kHz, respectively) is statistically insignificant. For a combined data set, we found distinct correlations between the universal law of drift rates and geometric mean frequencies of type II bursts is 0.96 with a power-law index of 1.66 $\pm$ 0.21 for a combined m to DH domain. We also found a good positive correlation between the slope of the DH type II bursts and space speed of CMEs for NIC-associated events(\emph{cc} = 0.71). But, this correlation is weak in the case of IC-associated events (\emph{cc} = 0.39). From these results we concluded that the shock speeds have been changed in the interplanetary medium due to the CME--CME interaction.\\[-5pt]

The average time lag between the start of m, DH type IIs, DH type III radio bursts, and the onset of SEP events for NIC-associated events is also lower than that of IC-associated events. These time lags studies suggest that NIC-associated SEP event particle enhancement happened at least 10 min before the IC-associated SEP events. It implies that the shock interaction and hence particle acceleration have been already progressive in the interplanetary space before the interaction of the leading edge of the CMEs.\\[-5pt]
 
We found that the peak intensity of SEPs for IC-associated events (559 pfu) is four times larger than that of NIC-associated SEP events (126 pfu). In Paper I, we reported positive correlations between peak intensity of SEPs and properties of parent solar eruptions (solar flare and CMEs) for NIC events. But such correlations were completely absent/weak in the case of IC events. Hence, from these results it is inferred that CME--CME interactions have played a vital role in the enhancements of the peak intensities of the SEP events. But we have noted the insignificant differences in the properties of DH type II radio bursts for NIC and IC-associated events might be due to the interaction of the narrower/lower speeds preceding CMEs.

\begin{acks}
We thank the referees for their strong commitment and useful constructive comments to improve the quality of the manuscript. We greatly acknowledge the data support provided by various online data centers of NOAA and NASA. We would like to thank the Wind/WAVES and RSTN spectrograph teams for providing the type II catalogs. The SOHO/LASCO CME catalog is generated and maintained at the CDAW Data Center by NASA and The Catholic University of America in cooperation with the Naval Research Laboratory. SOHO is a project of international cooperation between ESA and NASA. O. Prakash thanks the Chinese Academy of Sciences for providing General Financial Grant from the China Postdoctoral Science Foundation. 
\end{acks}





\section* {References}

\begin{enumerate}
    \item A. Vourlidas,P. Subramanian,K. P. Dere et al., Astrophys. J.534, 456 (2010) DOI:10.1086/308747.
    \item  N. Gopalswamy, Geo. Science Letts.3, 8 (2016)DOI: 10.1186/s40562-016-0039-2.
      \item N. Gopalswamy, S. Yashiro, G. Micha?ek et al., Astrophys. J. 572, L103 (2002) DOI:  10.1086/341601.
      \item O. Prakash, S. Umapathy, A. Shanmugaraju et al., Solar Phys. 281.765 (2012). DOI: 10.1007/s11207-010-9604-6
      \item V.V. Grechnev, A. M. Uralov, I. M. Chertok et al., Solar Phys, 273, 433 (2011). DOI: 10.1007/s11207-011-9780-z.
      \item R.-Y. Kwon, J. Zhang, O.Olmedo,Astrophys. J.794, 148 (2015). DOI: 10.1088/0004-637X/794/2/148
      \item A. P. Rouillard, I. Plotnikov, R. F. Pinto et al. Astrophys. J. 833, 45 (2016). DOI:  10.3847/1538-4357/833/1/45.
      \item N.Gopalswamy, E. Aguilar-Rodriguez, S.Yashiro et al. J. Geophys. Res. 110, A12S07 (2005). DOI: 10.1029/2005JA011158.
      \item N. Gopalswamy, H. Xie,  S. Yashiro et al., Space Sci Rev. 171, 23 (2012) . DOI: 10.1007/s11214-012-9890-4
   \item  J. P.Wild, Aus. J. Sci. Res. 3, 541 (1950). DOI: 10.1071/PH500541.
  \item  D. R. Nicholson, M. V. Goldman, P. Hoyng et al., Astrophys. J. 223, 605 (1978). DOI: 10.1086/156296
    \item  H. V. Cane, R. G. Stone,J. Fainberg et al., Geophys. Res. Lett.. 8, 1285 (1981) DOI: 10.1029/GL008i012p01285.
    \item  R. J. MacDowall, R. G. Stone, M. R. Kundu,Solar Phys. 111, 397 (1987) DOI:10.1007/BF00148528
  \item  H. V. Cane, W. C. Erickson, N. P. Prestage et al., J. Geophys. Res (Space Phys). 107, 1315 (2002), DOI: 10.1029/2001JA000320.
   \item  E. W. Cliver,  A. G. Ling, Astrophys. J. 690, 598 (2009). DOI: 10.1088/0004-637X/690/1/598.
    \item  L. M. Winter, K. Ledbetter, Astrophy. J. 809, 105 (2015). DOI: 10.1088/0004-637X/809/1/105.
 \item P. Pappa Kalaivani, A. Shanmugaraju, O. Prakash et al., Earth, Moon, and Planets, 123, 61 (2020). DOI: 10.1007/s11038-020-09533-9.
   \item  P. Pappa Kalaivani, O. Prakash, Li Feng et al., Adv. Space Res. 63, 3390 (2019) DOI : 10.1016/j.asr.2019.01.019
  \item  L. Ding, Y. Jiang, L. Zhao et al., Astrophys. J. 763, 30 (2013). DOI: 10.1088/0004-637X/763/1/30.
    \item  A. Shanmugaraju,S. Prasanna Subramanian,B.Vrsnak, Solar Phys, 289, 4621(2014) DOI:  10.1007/s11207-014-0591-x.
\item  O. Prakash, S. Umapathy, A. Shanmugaraju et al., Solar Phys. 258, 105 (2009). DOI: 10.1007/s11207-009-9396-8.
  \item  J. -L.Bougeret, M. L. Kaiser, P. J. Kellogg et al., Space Sci. Rev. 71, 231 (1995) DOI: 10.1007/BF00751331.
  \item M. H. Acuсa, K. W. Ogilvie,  D. N. Baker et al., Space Sci. Rev . 71, 5 (1995) DOI: 10.1007/BF00751323.
   \item  R. J. MacDowall, A. Lara, P. K. Manoharan et al., Geophys. Res. Lett. 30, 8018 (2003) DOI: 10.1029/2002GL016624.
 \item  G. Mann, A. Klassen, H.-T. Classen et al., Astron. Astrophys. Sup.119, 489 (1996) DOI:  1996A\&AS..119..489M
  \item  A. Shanmugaraju, Y.-J.Moon, M. Dryer et al. Solar Phys, 217, 301 (2003)DOI: 10.1023/B:SOLA.0000006902.89339.e4.
   \item  O. Prakash, Li, Feng, G. Michalek et al., Astrophys. Space Sci. 362, 56 (2017). DOI: 10.1007/s10509-017-3034-y.
  \item N.Gopalswamy, S. Yashiro, M. L. Kaiser et al. J. Geophys. Res. 106, 29219 (2001) DOI: 10.1029/2001JA000234
 \item  A. C. Umuhire, N. Gopalswamy, J. Uwamahoro et al., Solar Phys, 296, 27 (2021). DOI:10.1007/s11207-020-01743-8.
    \item  V. V. Lobzin, I. H.Cairns,  A. Zaslavsky,  J. Geophys. Res. 119, 742 ( 2014). DOI:10.1002/2013JA019008 
    \item  H. V. Cane, R. G. Stone,J. Fainberg et al., Geophys. Res. Letts. 8, 1285 (1981). DOI: 10.1029/GL008i012p01285.
\item  S. W. Kahler, J. Geophys. Res., 87, 3439 (1982). DOI: 10.1029/JA087iA05p03439.
  \item  D. Lario, A. Aran, R. Gуmez-Herrero et al., Astrophys. J.767, 41 (2013). DOI: 10.1088/0004-637X/767/1/41 
 \item A. P. Rouillard, N. R. Sheeley, A. Tylka et al., Astrophys. J. 752, 44 (2012).DOI:10.1088/0004-637X/752/1/44.
   \item  D. Lario, R.-Y. Kwon, A. Vourlidas et al., Astrophys. J. 819, 72 (2016). DOI: 10.3847/0004-637X/819/1/72.
\end{enumerate}    

\appendix   
\newgeometry{left=3cm, right=2cm}
\begin{table}
\tiny
\caption {Basic characteristics of m type II, DH type II and DH type III bursts for NIC and IC-associated SEP events}
\begin{threeparttable}
\begin{tabular}{|c|c|c|c|c|c|c|c|c|c|c|c|c|c|c|c|c|}

\hline
S.No & \shortstack{Date \\ (yyyy/mm/dd)} & \multicolumn{5}{c}{m type II data} & \multicolumn{6}{c}{DH type II data} & \multicolumn{3}{c}{DH type III data}\\
\cline{3-7} \cline{8-13} \cline{14-16}

	&				& \shortstack{RSTN\\ Station} & \shortstack{Start \\Time\\ (UT)} & \shortstack{End \\ Time \\(UT)} & \shortstack{Frequency\\ Range\\ (MHz)} & \shortstack{Shock \\ Speed \\ (km s$^{-1}$)} & \shortstack{Start \\ Time\\ (UT)} & \shortstack{End Date\\ (mm/dd)}&
		\shortstack{End \\ Time \\(UT)} & \shortstack{Frequency \\ range \\(kHz)} & \shortstack{Peak \\ Intensity\\ (sfu)} & \shortstack{Slope \\ (MHz  h$^{-1}$)} & \shortstack{Duration  \\ (min)} & \shortstack{Peak \\ Intensity\\ (sfu)} & \shortstack{Slope \\ (MHz  h$^{-1}$)}\\ \hline

\multicolumn {16}{c}{Non-Interacting CME-associated events}\\ \hline

1&1997/11/04&CULG&06:00&06:07&60-18X&1200&06:00&11/05&04:30&14000-100&4.97&0.38&29&6.83&14.76\\
2&1999/06/04&CULG&07:04&07:10&90-30&800&07:05&06/05&01:00&14000-60&4.36&0.85&30&7.13&23.47\\
3&2000/04/04&POTS&15:25&15:26&65-40X&----&15:45&04/04&16:00&14000-9000&2.81&0.16&25&6.88&25.63\\ 
4&2000/06/10&PALE&16:55&17:14&180-25&520&17:15&06/10&18:45&10000-1000&2.71&0.41&----&7.09&36.76\\
5&2001/04/02&CULG&21:52&21:57&55-28&800&22:05&04/03&02:30&14000-250&3.13&0.83&36&5.71&18.36\\
6&2001/09/15&SGMR&11:29&11:42&80-30&700&11:50&09/15&12:05&14000-6000&2.92&0.38&20&5.13&12.99\\
7&2001/10/22&HOLL&14:53&15:13&142-25&955&15:15&10/22&17:40&8000-1200&2.54&0.68&34&6.58&22.23\\
8&2001/11/17&LEAR&05:00&05:07&55-25&528&05:35&11/17&06:40&11000-1700&2.83&0.36&36&6.18&11.66\\
9&2001/11/22&PALE&20:22&20:47&180-25&890&20:50&11/22&22:23&8000-1000&2.71&0.64&33&6.51&21.75\\
10&2001/11/22&LEAR&22:31&22:41&116-25&459&22:40&11/24&02:30&14000-40&4.06&0.49&22&5.62&20.89\\
11&2002/01/14&CULG&06:08&06:11&90-43&1000&06:25&01/14&21:30&12000-100&2.55&0.45&26&6.10&13.75\\
12&2002/04/21&CULG&01:19&01:30&65-29X&500&01:30&04/21&24:00&10000-60&3.15&0.84&62&6.49&34.55\\
13&2002/08/14&LEAR&01:57&02:08&157-25&506&02:20&08/14&24:00&1000-30&3.51&0.94&44&6.75&28.95\\
14&2002/09/05&SVTO&16:35&16:49&180-32&679&16:55&09/07&16:22&14000-30&5.08&0.84&15&6.00&12.41\\
15&2003/05/28&PALE&00:26&00:33&180 -25&392&01:00&05/29&00:30&1000-200&3.76&0.31&20&6.41&12.81\\
16&2003/06/17&CULG&22:48&22:58&90-18&1000&22:50&06/18&05:30&10000-200&3.54&0.98&11&5.95&14.33\\
17&2004/09/12&CULG&0:23&00:29&57-30&900&00:45&09/13&21:00&14000-40&4.39&0.56&32&6.02&17.19\\
18&2004/11/09&PALE&17:18&17:21&56-25&1866&17:35&11/09&21:28&14000-600&2.83&0.38&26&6.71&18.28\\
19&2004/11/10&LEAR&02:07&02:40&180-25&1023&02:25&11/10&03:40&14000-1000&3.51&1.26&29&6.56&18.31\\
20&2005/01/15&CULG&22:34&22:42&45-20&1900&23:00&01/17&00:00&3000-40&4.21&1.81&28&5.23&19.27\\
21&2005/05/13&SVTO&16:41&16:52&81-25&1349&17:00&05/15&02:10&5000-40&3.26&0.64&24&7.15&23.24\\
22&2011/03/07&SGMR&19:54&20:05&120-18X&----&20:00&03/08&08:30&16000-200&4.09&0.63&30&6.24&19.30\\
23&2011/08/04&CULG&03:53&04:10&90-18&600&04:15&08/05&17:00&13000-60&3.13&0.84&40&6.09&19.51\\
24&2011/08/09&LEAR&08:08&08:17&180-25&1200&08:20&08/09&08:35&16000-4000&3.24&0.94&31&6.28&32.68\\
25&2011/09/22&SVTO&10:39&10:46&85-25&968&11:05&09/22&24:00&14000-70&4.27&0.62&33&6.28&18.45\\
26&2012/01/27&SGMR&18:10&18:21&90-18X&790&18:30&01/28&04:45&16000-150&3.90&0.53&33&5.60&19.04\\
27&2012/03/07&LEAR&00:17&00:31&90-18&2273&01:00&03/08&19:00&16000-30&3.96&0.57&29&7.85&16.58\\
28&2012/03/13&SGMR&17:17&17:29&180-25&1060&17:35&03/13&24:00&16000-200&4.14&1.67&35&6.66&40.74\\
29&2012/07/06&CULG&2310&23:18&90-23&1500&23:10&07/07&03:40&16000-300&4.60&0.20&21&6.62&33.52\\
30&2012/08/31&SGMR&19:42&19:55&62-25X&856&20:00&08/31&23:45&16000-400&3.95&0.62&45&5.74&15.73\\
31&2012/09/27&CULG&23:44&23:52&30-18X&700&23:55&09/28&10:1&16000-250&4.64&0.33&45&5.61&17.70\\
32&2013/04/11&CULG&07:02&07:09&80-23&1000&07:10&04/11&15:00&10000-200&4.52&0.38&35&7.21&19.38\\
33&2014/02/20&CULG&07:46&07:51&75-25&900&08:05&02/20&08:29&12000-7700&2.33&0.97&13&6.30&31.22\\
34&2014/02/25&LEAR&00:56&01:00&90-18&930&00:56&02/25&11:28&14000-100&4.31&0.45&31&7.16&14.81\\
35&2014/04/18&SVTO&12:55&13:04&80-25X&860&13:05&04/18&22:50&14000-150&3.70&0.58&31&6.12&20.19\\ \hline
\multicolumn {16}{c}{Interacting CME-associated events}\\ \hline
36&1998/04/20&SVTO&09:56&10:02&51-35&2000&10:25&04/22&06:00&10000-35&4.25&0.57&13&6.70&17.15\\
37&1998/05/09&CULG&03:26&03:29&75-23&1500&03:35&05/09&10:00&9000-400&3.10&0.66&33&5.53&13.43\\
38&2000/06/06&HOLL&15:23&15:27&146-25&1189&15:20&06/08&09:00&14000-40&4.59&0.79&41&7.00&10.52\\
39&2000/09/12&SGMR&11:42&11:47&60-30&1800&12:00&09/13&12:20&14000-60&3.84&0.37&39&6.42&13.94\\
40&2000/10/16&SVTO&07:08&07:19&56-25&806&07:10&10/16&08:00&14000-1000&2.74&0.54&13&5.35&14.42\\
41&2000/11/08&CULG&23:16&23:22&200-100&----&23:20&11/09&12:00&4000-200&3.26&0.67&48&7.47&34.43\\
42&2000/11/24&HOLL&15:07&15:14&180-38&1200&15:25&11/24&22:00&14000-200&4.01&0.72&33&6.07&16.25\\
43&2001/04/10&CULG&05:13&05:17&70-20&2000&05:24&04/10&24:00&14000-100&3.72&0.92&40&7.57&25.24\\
44&2001/10/19&HOLL&16:24&16:42&180-25&738&16:45&10/21&16:40&14000-30&4.89&0.62&26&7.16&13.45\\
45&2002/07/20&PALE&21:07&21:29&180-25&515&21:30&07/20&22:20&10000-2000&***&----&----&----&----\\
46&2002/11/09&SGMR&13:22&13:25&65-30&700&13:20&11/10&03:00&14000-100&3.18&0.55&31&6.52&16.50\\
47&2003/05/31&CULG&02:23&02:35&90-18&800&03:00&05/31&08:00&1000-150&2.51&0.97&24&6.75&31.59\\
48&2003/10/26&PALE&17:35&17:43&100-25&940&17:45&10/26&19:40&14000-1500&3.09&0.31&21&5.62&35.52\\
49&2003/11/04&HOLL&19:42&19:50&180-25&1074&20:00&11/04&24:00&10000-200&4.72&1.31&37&7.31&20.89\\
50&2004/07/25&SVTO&15:21&15:26&81-25&898&15:00&07/26&22:25&1000-28&3.72&0.98&5&5.86&14.02\\
51&2004/11/07&SGMR&15:59&16:16&180-25&697&16:25&11/08&20:00&14000-60&4.33&1.26&42&5.72&13.15\\
52&2006/12/14&LEAR&22:09&22:13&180-25&1277&22:30&12/14&23:40&14000-1500&2.61&0.42&23&6.51&29.72\\
53&2011/06/07&CULG&06:26&06:40&70-15&850&06:45&06/07&18:00&16000-250&3.98&0.54&31&7.19&30.80\\
54&2012/07/08&SVTO&16:30&16:43&150-25&1210&16:35&07/08&22:00&16000-300&2.88&0.82&38&4.89&11.86\\
55&2012/07/12&SGMR&16:27&16:38&50-25X&900&16:45&07/13&09:00&14000-250&3.45&0.45&38&5.27&18.00\\
56&2012/07/19&CULG&05:24&05:38&55-18&1100&05:30&07/19&06:20&5000-600&3.61&1.91&15&6.40&22.97\\
57&2013/05/22&SVTO&12:59&13:03&40-25X&1055&13:10&05/24&06:00&16000-150&6.37&0.72&35&5.98&14.77\\
58&2014/01/07&SGMR&18:17&18:23&45-25&1210&18:33&01/08&21:00&14000-60&3.72&0.83&49&6.69&28.25\\ \hline

\end{tabular}

\begin{tablenotes}
\item{In Column 6 the letter X indicates with a number that the emission continued beyond the instrument range; ---- Means that unavailable of the corresponding data.}
\end{tablenotes}

\end{threeparttable}
\end{table}


\restoregeometry
\end{article}

\end{document}